\documentclass[sigconf, authorversion]{acmart}

\usepackage[utf8]{inputenc}
\usepackage{rotating}
\usepackage{listings}
\usepackage{nicefrac}
\usepackage{color,colortbl}

\usepackage{siunitx}
\DeclareSIUnit{\billion}{B}

\usepackage{makecell}

\usepackage{multirow}
\usepackage{subcaption}
\definecolor{lightgray}{gray}{0.9}

\usepackage{stfloats}

\usepackage[ruled]{algorithm2e}
\definecolor{color-best}{rgb}{1, 0, 0}
\definecolor{color-best-dark}{rgb}{0.5, 0.8, 0.4}
\definecolor{color-best-light}{rgb}{0.6, 0.9, 0.5}
\definecolor{color-best-right-dark}{rgb}{0.65, 0.95, 0.55}
\definecolor{color-best-right-light}{rgb}{0.7, 1, 0.6}
\definecolor{color-coarse}{rgb}{0, 0.5, 0}
\definecolor{color-outlier-dark}{rgb}{0.8, 0.5, 0.4}
\definecolor{color-outlier-light}{rgb}{0.9, 0.6, 0.5}

\newcommand{\bestcell}{%
  \ifodd\rownum
    \cellcolor{color-best-dark}
  \else
    \cellcolor{color-best-light}
  \fi
}
\newcommand{\bestcellv}{%
  \ifodd\rownum
    \cellcolor{color-best-light}
  \else
    \cellcolor{color-best-dark}
  \fi
}
\newcommand{\outliercell}{%
  \ifodd\rownum
    \cellcolor{color-outlier-dark}
  \else
    \cellcolor{color-outlier-light}
  \fi
}
\newcommand{\bestcellright}[1]{%
  \ifodd\rownum
    \cellcolor{color-best-right-light}
  \else
    \cellcolor{color-best-right-light}
  \fi
  #1 \hspace{-2pt}\makebox[0pt][r]{\makebox[1pt][l]{\raisebox{-2pt}{\rule{2pt}{0.5pt}\rule{0.5pt}{2pt}}}}%
}

\usepackage{xspace}

\usepackage{enumitem}

\usepackage[toc, acronym, xindy, shortcuts]{glossaries}

\newacronym{asan}{\mbox{ASan}}{AddressSanitizer}
\newacronym{dbcv}{\mbox{DBCV}}{Density Based Cluster Validity}
\newacronym{dbscan}{\mbox{DBSCAN}}{Density-Based Spatial Clustering of Applications with Noise}
\glsunset{dbscan}
\newacronym{hdbscan}{\mbox{HDBSCAN}}{Hierarchical Density-Based Spatial Clustering of Applications with Noise}
\newacronym{jvm}{\mbox{JVM}}{Java Virtual Machine}
\newacronym{llm}{\mbox{LLM}}{large language model}
\newacronym{mteb}{\mbox{MTEB}}{Massive Text Embedding Benchmark}
\newacronym{sci}{\mbox{SCI}}{sample crashing input}
\glsdisablehyper

\newcommand{\name}{GPTrace\xspace}
\newcommand{\repourl}{\url{https://github.com/Fraunhofer-AISEC/gptrace-artifacts}}
\newcommand{\tembed}{text-embedding-3-large\xspace}
\newcommand{\tembedshort}{text-emb.-3-large\xspace}
\newcommand{\nvembed}{NV-Embed-v2\xspace}
\newcommand{\stella}{stella\_en\_1.5B\_v5\xspace}
\newcommand{\cpp}{C/C\texttt{++}\xspace}
\newcommand{\aflpp}{AFL\texttt{++}\xspace}
\usepackage{mathtools}
\DeclarePairedDelimiter\abs{\lvert}{\rvert}
\newcommand{\norm}[1]{\left\lVert#1\right\rVert}

\newcommand*{\eg}{e.g.,\@\xspace}
\newcommand*{\ie}{i.e.,\@\xspace}

\usepackage{cleveref}
\crefformat{footnote}{#2\footnotemark[#1]#3}

\copyrightyear{2026}
\acmYear{2026}
\setcopyright{cc}
\setcctype{by}
\acmConference[ICSE '26]{2026 IEEE/ACM 48th International Conference on Software Engineering}{April 12--18, 2026}{Rio de Janeiro, Brazil}
\acmBooktitle{2026 IEEE/ACM 48th International Conference on Software Engineering (ICSE '26), April 12--18, 2026, Rio de Janeiro, Brazil}
\acmPrice{}
\acmDOI{10.1145/3744916.3773161}
\acmISBN{979-8-4007-2025-3/2026/04}


\begin{document}

\title{GPTrace: Effective Crash Deduplication Using LLM Embeddings}

\author{Patrick Herter}
\orcid{0009-0004-7829-6179}
\affiliation{%
  \institution{Fraunhofer AISEC}
  \city{Garching near Munich}
  \country{Germany}
}
\affiliation{%
  \institution{Technical University of Munich}
  \city{Munich}
  \country{Germany}
}
\email{patrick.herter@aisec.fraunhofer.de}

\author{Vincent Ahlrichs}
\orcid{0000-0002-0893-0822}
\affiliation{%
  \institution{Fraunhofer AISEC}
  \city{Garching near Munich}
  \country{Germany}
}
\affiliation{%
  \institution{Technical University of Munich}
  \city{Munich}
  \country{Germany}
}
\email{vincent.ahlrichs@aisec.fraunhofer.de}

\author{Ridvan Açilan}
\orcid{0009-0008-7344-2770}
\affiliation{%
  \institution{Technical University of Munich}
  \city{Munich}
  \country{Germany}
}
\email{ridvan-acilan@hotmail.de}

\author{Julian Horsch}
\orcid{0000-0001-9018-7048}
\affiliation{%
  \institution{Fraunhofer AISEC}
  \city{Garching near Munich}
  \country{Germany}
}
\email{julian.horsch@aisec.fraunhofer.de}

\begin{abstract}
  Fuzzing is a highly effective method for uncovering software vulnerabilities,
but analyzing the resulting data typically requires substantial manual effort.
This is amplified by the fact that fuzzing campaigns often find a large number
of crashing inputs, many of which share the same underlying bug.
Crash deduplication is the task of finding such duplicate crashing inputs and
thereby reducing the data that needs to be examined.
Many existing deduplication approaches rely on comparing stack traces or
other information that is collected when a program crashes.
Although various metrics for measuring the similarity of such pieces of
information have been proposed, many do not yield satisfactory deduplication
results.
In this work, we present \name, a deduplication workflow that leverages a
\glsentrylong{llm} to evaluate the similarity of various data sources associated
with crashes by computing embedding vectors and supplying those as input to a
clustering algorithm.
We evaluate our approach on over \num{300000} crashing inputs belonging to
50 ground truth labels from 14 different targets.
The deduplication results produced by \name show a noticeable improvement over
hand-crafted stack trace comparison methods and even more complex
state-of-the-art approaches that are less flexible.
\end{abstract}


\maketitle

\section{Introduction}

Fuzzing is one of the most popular and effective methods for testing software
and finding security vulnerabilities therein.
A fuzzing tool generates a large amount of test inputs either by randomized
mutation of a valid seed corpus or by deriving them from a specification format.
Inputs that cause a crash of the target software are collected along with
some related information about the crash.
Using these \emph{\glspl{sci}}, developers try to analyze their software's
behavior, find the underlying bugs, and fix them.
However, fuzzing campaigns tend to produce thousands of \glspl{sci}, among which
many trigger the same underlying bug.
This results in significant resources being spent on organizing, assigning, and
analyzing issues, only to find that many are duplicates.

Deduplication tools aim to remedy this by attempting to group \glspl{sci} that
likely trigger the same underlying bug.
A common deduplication approach (employed, \eg by Crashwalk~\cite{crashwalk},
TraceSim~\cite{tracesim} and ReBucket~\cite{rebucket}) is to collect stack
traces when the \glspl{sci} trigger a crash and define a suitable notion of
similarity of these stack traces.
Similarity scores can then be used to decide which \glspl{sci} should
be grouped together.
Many approaches that deduplicate based on stack trace analysis devise
hand-crafted algorithms that produce similarity scores based on the syntactic
properties of stack traces.
These are often inflexible and fail to capture the full information that
is contained in stack traces, thus yielding unsatisfactory results.

With the emergence of \glspl{llm} we have a new tool at our disposal to process
textual input in various ways.
Internally, \glspl{llm} use vectors of numbers to represent input queries.
These \emph{embeddings} ought to capture the semantic meaning of the
input such that distances between embedding vectors reflect the similarity of
the corresponding inputs.
Our proposed deduplication method, \name, computes \gls{llm} embeddings of stack
traces and \gls{asan}~\cite{asan} reports.
Using standard distance functions on numerical vectors, \eg the Euclidean
metric, we can run clustering algorithms, such as
\gls{hdbscan}~\cite{hdbscan, hdbscan_2, hdbscan_impl}, on the embedding vectors
and thereby obtain a grouping of our \glspl{sci}.

By leveraging \gls{llm} embeddings and the contained knowledge gained
from large amounts of data that were used during training of the \gls{llm},
\name can extract more information from the supplied stack traces and
\gls{asan} reports, and provide more robust similarity assessments than existing
fixed, hand-crafted stack trace analysis methods.
This manifests itself in accurate deduplication results that improve on those of
existing stack trace deduplication approaches and even more sophisticated
approaches that require more detailed execution information, such as
Igor~\cite{igor} or DeFault~\cite{default}.
Moreover, \name's flexible design enables integration of different kinds of
crash data on a per-target basis and can be adapted for use with targets
written in different programming languages.

In summary, our concrete contributions are the following:
\begin{itemize}
    \item A concept for leveraging \gls{llm} embeddings for the purpose of
          crash deduplication.
    \item A prototype implementation of \name written in Python for
          deduplication of \glspl{sci} for \cpp programs, which we provide as
          open source.\footnote{
              \label{note:artifacts-url}
              \repourl
          }
    \item An extensive evaluation of different variants of \name as well as
          comparisons to existing deduplication approaches.
          Our artifacts for the evaluation are also publicly
          available.\footref{note:artifacts-url}
\end{itemize}
In the rest of this paper, we present the design of \name (\autoref{sec:design})
in detail and then discuss some notable points about our prototype
implementation (\autoref{sec:implementation}).
Next, we present the results of our evaluation (\autoref{sec:evaluation}),
including a comparison of \name to previous crash deduplication methods.
Finally, we talk about related work (\autoref{sec:related-work}) before drawing
a conclusion (\autoref{sec:conclusion}).

\section{Design}
\label{sec:design}

\begin{figure*}[ht]
    \centering
    \resizebox{\textwidth}{!}{
        \includegraphics{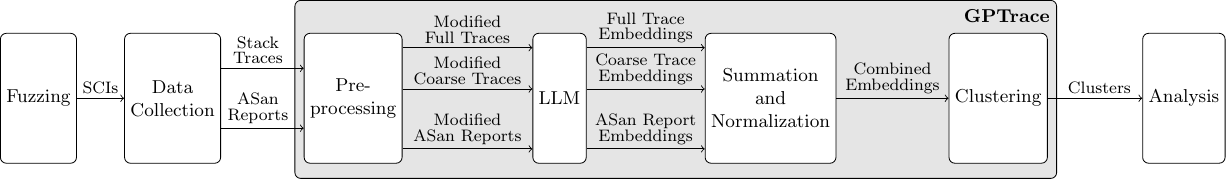}
    }
    \Description{
        Overview of the \name workflow.
        The design is described in the text.
    }
    \caption{Overview of the \name workflow.}
    \label{fig:design}
\end{figure*}

In this section, we describe the design of \name, an overview of which is
shown in \autoref{fig:design}.
For each \gls{sci} returned by a fuzzing tool, \name processes
associated data that contains information about the crashing behavior of that
\gls{sci}.
Valuable information is carried by stack traces and \gls{asan} reports
captured during the execution of the target software with the \gls{sci}.
As a first step, these data sources undergo a preprocessing phase, during which
several modifications take place.
Next, \name collects for each of these modified data sources the associated
\gls{llm} embedding.
We combine these individual embedding vectors for each \gls{sci}
and then employ a clustering process that yields as a result our final bucketing
of the \glspl{sci}.

\subsection{Preprocessing}
\label{sec:design/preprocessing}

\begin{figure*}
    \begin{tabular*}{\textwidth}{c@{\extracolsep{\fill}}c}
        \captionsetup[subfigure]{aboveskip=3pt}
        \multirowcell{2}[135pt]{
            \begin{subfigure}{0.47\textwidth}
\begin{lstlisting}[
                    basicstyle=\ttfamily\footnotesize, frame=single,
                    breaklines=true, escapechar=|
                ]
#0  0xf7fc4579 in __kernel_vsyscall |\color{color-coarse}()|
#1  0xf7799eb7 in ?? |\color{color-coarse}()| from /lib32/libc.so.6
#2  0xf77494c5 in raise |\color{color-coarse}()| from /lib32/libc.so.6
#3  0xf77323ac in abort |\color{color-coarse}()| from /lib32/libc.so.6
#4  0x0815f276 in __sanitizer::Abort|\color{color-coarse}() ()|
#5  0x0815d377 in __sanitizer::Die|\color{color-coarse}() ()|
#6  0x08138f8f in __asan::ReportGenericError|\color{color-coarse}(unsigned long,|
               |\color{color-coarse}unsigned long, unsigned long, unsigned long,|
               |\color{color-coarse}bool, unsigned long, unsigned int, bool) ()|
#7  0x0814158a in __asan_report_load1 |\color{color-coarse}()|
#8  0xf7e8bfbb in tt_cmap10_char_index |\color{color-coarse}(cmap=0xf4c00940,|
               |\color{color-coarse}char\_code=64)|
               at /freetype-2.5.3/src/sfnt/ttcmap.c:1931
#9  0xf7cdc271 in FT_Get_Char_Index |\color{color-coarse}(face=0xf4203c80,|
               |\color{color-coarse}charcode=64)|
               at /freetype-2.5.3/src/base/ftobjs.c:3347
#10 0x0817a5f3 in OutlinePrinter::LoadGlyph |\color{color-coarse}(this=|
               |\color{color-coarse}<optimized out>, symbol=<optimized out>)|
               at char2svg.cpp:246
#11 OutlinePrinter::Run |\color{color-coarse}(this=<optimized out>, symbol=|
               |\color{color-coarse}<optimized out>)| at char2svg.cpp:217
#12 0x0817d6f0 in main |\color{color-coarse}(argc=<optimized out>, argv=|
               |\color{color-coarse}<optimized out>)| at char2svg.cpp:486
\end{lstlisting}
                \caption{
                    Full stack trace. The parts that are removed to obtain
                    the coarse stack trace in \autoref{fig:coarse-trace} are
                    colored in green.
                }
            \end{subfigure}
        }
        &
        \captionsetup[subfigure]{aboveskip=3pt, belowskip=8pt}
        \begin{subfigure}{0.47\textwidth}
\begin{lstlisting}[
                basicstyle=\ttfamily\footnotesize, frame=single,
                breaklines=true
            ]
#0  0xf7fc4579 in __kernel_vsyscall
#1  0xf7799eb7 in ??  from /lib32/libc.so.6
#2  0xf77494c5 in raise  from /lib32/libc.so.6
#3  0xf77323ac in abort  from /lib32/libc.so.6
#4  0x0815f276 in __sanitizer::Abort
#5  0x0815d377 in __sanitizer::Die
#6  0x08138f8f in __asan::ReportGenericError
#7  0x0814158a in __asan_report_load1
#8  0xf7e8bfbb in tt_cmap10_char_index
               at /freetype-2.5.3/src/sfnt/ttcmap.c:1931
#9  0xf7cdc271 in FT_Get_Char_Index
               at /freetype-2.5.3/src/base/ftobjs.c:3347
#10 0x0817a5f3 in OutlinePrinter::LoadGlyph
               at char2svg.cpp:246
#11 OutlinePrinter::Run at char2svg.cpp:217
#12 0x0817d6f0 in main at char2svg.cpp:486
\end{lstlisting}
            \caption{
                Coarse stack trace with all function arguments removed.
            }
            \label{fig:coarse-trace}
        \end{subfigure}
        \\
        &
        \captionsetup[subfigure]{aboveskip=3pt, belowskip=-2pt}
        \begin{subfigure}{0.465\textwidth}
\begin{lstlisting}[
                basicstyle=\ttfamily\footnotesize, frame=single,
                breaklines=true
            ]
ERROR: AddressSanitizer: heap-buffer-overflow
       on address 0xf4e0382a
       at pc 0xf7e8bfbb bp 0xffffd8f8 sp 0xffffd8ec
READ of size 1 at 0xf4e0382a thread T0
Address 0xf4e0382a is a wild pointer.
SUMMARY: AddressSanitizer: heap-buffer-overflow
         /freetype-2.5.3/src/sfnt/ttcmap.c:1931:16
         in tt_cmap10_char_index
\end{lstlisting}
            \caption{
                \gls{asan} report with removed stack trace, shadow map and
                legend.
            }
        \end{subfigure}
    \end{tabular*}
    \Description{
        Examples of data sources used by \name.
        These were produced by the target char2svg.
    }
    \caption{
        Examples of data sources used by \name.
        These were produced by the target char2svg.
    }
    \label{fig:example-data-sources}
\end{figure*}

Before computing embeddings, \name applies modifications to each of the input
data sources.
For every \gls{sci}, we use two versions of the same stack trace.
We found that for some targets, stack traces can contain thousands of copies of
the same frame.
To reduce the size of the processed data, we keep only the topmost copy and
discard all others from both versions of the stack trace.
In addition to this, the second version of the stack trace is heavily shortened
by removing all function arguments from it.
This results in stack traces that only contain coarse but crucial information,
which is thereby given more weight in the following steps.

\gls{asan} reports also contain a copy of the stack trace.
However, since \name tracks this information separately, the traces are removed
from the reports to give more weight to the crucial \gls{asan} information.
A simpler approach would be to not process the stack traces separately and
instead to solely rely on \gls{asan} reports with the contained stack traces.
However, we show in \autoref{sec:evaluation/data-sources} that processing these
pieces of data separately improves the results.

To further increase the signal-to-noise ratio, we also remove the shadow byte
map and legend that \gls{asan} reports contain.
Examples of full and coarse stack traces as well as a modified \gls{asan}
report produced this way are shown in \autoref{fig:example-data-sources}.
Before moving on to the next step, \name searches the processed data for
duplicate entries for which all modified data sources are identical, and
discards them.

\subsection{Embedding}
\label{sec:design/embedding}

After preparing the various data sources, \name uses an \gls{llm} to compute
embeddings.
These embeddings are vectors of numbers that are used by the \gls{llm} for its
internal processing.
As such, they capture the semantic meaning of the input from the point of
view of the \gls{llm}.
The distance between embedding vectors of different inputs determines how
similar these inputs are in the perception of the \gls{llm}.
This is the key principle that \name leverages.

In more detail, \name obtains a separate embedding vector $v_{d, s}$ for each
data source $d \in D \coloneqq \{\text{Full Traces},\allowbreak
    \text{Coarse Traces},\allowbreak \text{\gls{asan} Reports}\}$ and each
$s \in \{\text{\glspl{sci}}\}$.
We then combine, for a given $s \in \{\text{\glspl{sci}}\}$, the various
$v_{d, s}$, $d \in D$, into a single vector as follows:
\begin{displaymath}
    v_s \coloneqq \sum_{d \in D} \frac{v_{d, s}}{\norm{v_{d, s}}}.
\end{displaymath}
Here, $\norm{\cdot}$ denotes the Euclidean norm, and the normalization ensures
that one vector does not dominate the others in the sum due to large entries.
In practice, \glspl{llm} often produce embedding vectors that are already
normalized.
We decided to sum the vectors instead of concatenating them in order to enhance
features that are shared across data sources:
Each dimension of the resulting embedding vectors corresponds to some feature
that the \gls{llm} associates with the input, independent of the data source.
Summing the vectors amplifies a feature in the resulting vector if it is present
in multiple data sources, while concatenating would keep them separate.
Additionally, concatenation would produce higher-dimensional vectors, which
tends to degrade the performance of clustering algorithms.
\enlargethispage{-0.05\baselineskip}

Quantifying the similarity of the obtained vectors during the clustering phase
requires a distance function.
While the cosine distance is commonly used for \gls{llm} embedding vectors, it
does not obey the triangle inequality.
Several computations can be accelerated by using a distance function that does
obey the triangle inequality, such as the Euclidean metric.
When restricting to vectors that have a Euclidean norm of $1$, the orderings of
comparisons for the cosine distance and the Euclidean distance are the same.
Thus, we normalize the final vectors with respect to the Euclidean metric
and then use the Euclidean distance as our distance function in the subsequent
steps.
These combined, normalized vectors are now passed on to the clustering stage.

\subsection{Clustering}
\label{sec:design/clustering}

Clustering algorithms generally provide parameters that
control how coarse or fine-grained the resulting clusters are, requiring
us to search for values that yield satisfactory results.
This makes clustering algorithms that require the number of resulting clusters
as a parameter, like $k$-means~\cite{kmeans}, unsuitable for our use case, since
this value can range from $1$ to the number of \glspl{sci} supplied to \name,
rendering a search impractical when analyzing a large number of crashing inputs.
Moreover, we will see in \autoref{sec:evaluation/clustering} that $k$-means
does not produce optimal results for our use case.

We found density-based clustering algorithms to provide good results for our
application and since the arrangement and density of the embedding vectors is
unknown and may vary both within a target as well as across different targets,
we found a variable density clustering using
\gls{hdbscan}~\cite{hdbscan, hdbscan_2, hdbscan_impl} to be more suitable than
fixed density algorithms like \gls{dbscan}~\cite{dbscan}.
\gls{hdbscan} creates a cluster hierarchy and evaluates cluster stability over
varying distance thresholds.
It takes the parameters $m_\text{clSize}$, which determines the minimum number
of points a cluster can have, and $m_\text{pts}$, which influences how
conservative the clustering will be.
Large values of $m_\text{pts}$ will result in many points declared as noise and
only produce clusters in very dense areas.
As there might well be only a single \gls{sci} for some bugs, we set
these parameters as low as possible ($m_\text{clSize} = 2$ and
$m_\text{pts} = 1$).
Since this configuration is prone to creating a lot of
micro-clusters, even in regions of high density, we use the hybrid approach by
\citeauthor{hybrid_hdbscan}~\cite{hybrid_hdbscan}, which combines \gls{dbscan}
and \gls{hdbscan} by using fixed density \gls{dbscan} clusters below a
distance threshold $\varepsilon$, which is provided as a parameter.

\begin{algorithm}[t]
    \caption{
        \name search algorithm for a good clustering. Here, \emph{min\_dist}
        and \emph{max\_dist} are the minimum and maximum distances between
        data points respectively.
        \emph{num\_steps} specifies an upper limit on the number of
        iterations.
    }
    \label{algo:cluster-search}
    \DontPrintSemicolon
    \SetArgSty{relax}
    \SetFuncSty{relax}
    \SetFuncArgSty{relax}
    \SetKwFunction{cls}{cluster\_search}
    \SetKwProg{proc}{procedure}{}{}
    \proc{\cls{\text{min\_dist}, \text{max\_dist}, \text{num\_steps}}}{

    $\text{queue} \gets [(\text{min\_dist}, \text{max\_dist})]$\;
    $\text{clusterings} \gets []$\;
    $\text{steps} \gets 0$\;

    \While{$\operatorname{not\_empty}(\text{queue})$ and $\text{steps} < \text{num\_steps}$}{
    $(\text{eps\_start}, \text{eps\_end}) \gets \operatorname{dequeue}(\text{queue})$\;
    $\text{clustering\_start} \gets \operatorname{hdbscan}(\text{eps\_start})$\;
    $\text{clustering\_end} \gets \operatorname{hdbscan}(\text{eps\_end})$\;

    $\operatorname{append}(\text{clusterings}, \text{clustering\_start})$\;
    $\operatorname{append}(\text{clusterings}, \text{clustering\_end})$\;

    \If{$\text{clustering\_start} \ne \text{clustering\_end}$}{
        $\text{eps\_mid} \gets \frac{\text{eps\_start} + \text{eps\_end}}{2}$\;
        $\text{step} \gets \frac{\text{eps\_mid} - \text{eps\_start}}{\text{num\_steps}}$\;
        $\operatorname{enqueue}(\text{queue}, (\text{eps\_start} + \text{step}, \text{eps\_mid}))$\;
        $\operatorname{enqueue}(\text{queue}, (\text{eps\_mid} + \text{step}, \text{eps\_end} - \text{step}))$\;
        }

        $\text{steps} \gets \text{steps} + 1$\;
        }

        \Return{$\operatorname{choose\_best\_clustering}(\text{clusterings})$}
    }
\end{algorithm}
In this setup, $\varepsilon$ is the main parameter that greatly
influences the clustering result.
\name tries several values for $\varepsilon$ and chooses the one that produces
the best clustering based on a strategy that we describe below.
To achieve a time-efficient search for $\varepsilon$, we devise
\autoref{algo:cluster-search}.
Here, we limit the search interval to the range of distance values that actually
occur in the dataset, skipping values that do not produce additional
clusterings.
Furthermore, we employ a divide-and-conquer approach to skip over
$\varepsilon$-ranges that leave the clustering unchanged.
Since $\varepsilon$ only influences the final selection of the clustering from
the cluster hierarchy, the computation of the minimum spanning tree of the
mutual reachability graph can be done once, independently of $\varepsilon$.
This is the most time-consuming part of \gls{hdbscan} and caching its result
greatly improves the run time of the individual \gls{hdbscan} runs and thereby
of \autoref{algo:cluster-search}.

The function \emph{choose\_best\_clustering} used in
\autoref{algo:cluster-search} quantifies the quality of each clustering
produced by \gls{hdbscan} and chooses the best one based on the following
three metrics:
\begin{description}[style=unboxed]
    \item[\gls{dbcv}:]
        \gls{dbcv}~\cite{dbcv} is a relative metric that uses values between
        $-1$ and $1$ to measure the quality of a clustering based on computing
        minimal density scores within clusters and maximal density scores
        between pairs of clusters.
        \gls{dbcv} is particularly suitable for quantifying the quality of
        density-based clusterings.
    \item[Persistence:]
        The persistence of a clustering is a value between $0$ and $1$ that
        measures the distance range through which the clustering persists in
        the cluster hierarchy computed by \gls{hdbscan}.
        We have found that clusterings with larger persistence values more
        closely resemble the ground truth.
    \item[Number of clusters:]
        We count each point that \gls{hdbscan} classifies as noise as a separate
        cluster.
        Although the number of clusters does not allow conclusions about the
        quality of the clustering itself, we prioritize clusterings with a
        lower number of clusters, as additional clusters result in potential
        manual work for humans.
        As fuzzing and bug fixing are iterative processes, a bug overlooked due
        to coarser clustering is likely to resurface in subsequent bug fixing
        rounds.
\end{description}
Based on these metrics, \emph{choose\_best\_clustering} selects the best
clustering from all candidates using the following strategy:
\begin{enumerate}
    \item Record the largest \gls{dbcv} score that occurs among all clusterings
          and then choose at most ten clusterings with the largest \gls{dbcv}
          scores that deviate from the maximum re\-corded score by at most
          \qty{20}{\%}.
    \item Among the resulting clusterings, record the
          largest persistence value and then choose all clusterings with
          persistence values that deviate from the maximum by
          at most \qty{20}{\%}.
    \item Among the resulting clusterings, choose the
          one with the lowest number of clusters.
          Since the first two steps limit the deviation from the maximum
          \gls{dbcv} score and persistence, only clusterings with acceptable
          accuracy will reach this step.
          Thus, we select the clustering that minimizes the subsequent human
          analysis effort, as explained above.
\end{enumerate}
The final result of \name is a clustering of the supplied \glspl{sci} where
each cluster is a group of \glspl{sci} that \name suspects to have the same
underlying bug.

\section{Implementation}
\label{sec:implementation}

We implemented a prototype of \name in Python 3.11.
Although the general idea of \name can be applied to deduplication of
\glspl{sci} for programs written in any language, our prototype implementation
targets \cpp programs.

During development, we used OpenAI's \tembed model~\cite{openai_embedding}
for computing embeddings and we will discuss some specific adjustments we made
in our implementation using the OpenAI Python library~\cite{openai_python}.
However, our prototype also supports
SentenceTransformer~\cite{sentence_transformer} models, \eg from Hugging
Face~\cite{huggingface}.

The embedding vectors that \tembed returns have a length of 3072.
However, many clustering algorithms, including \gls{hdbscan}, struggle with
very-high dimensional data.
Many state-of-the-art \glspl{llm}, including \tembed, are trained
using Matryoshka Representation Learning~\cite{matryoshka}, a technique that
produces embeddings where the first components of the vectors contain the most
crucial information while the later components become more fine-grained.
This allows us to truncate our embedding vectors to 64 dimensions without
incurring significant loss of information.
As these lower-dimensional vectors can be more easily handled by our clustering
algorithm, this results in an overall gain in deduplication quality with a
positive side-effect on run times.

Deduplication tools need to deal with large numbers of \glspl{sci} and
sending individual embedding requests for the data of individual \glspl{sci} to
the OpenAI API results in slow run times and rate limiting.
To remedy this, we make use of OpenAI batch processing, which allows creating
asynchronous jobs that consist of a large number of requests.
To avoid computing embeddings multiple times during consecutive runs of \name,
we store the embeddings obtained by the \gls{llm} and label them using the hash
of the data they belong to.

To achieve a fast clustering phase, we use the Python hdbscan
library~\cite{hdbscan_impl}, which implements the hybrid clustering approach
as well as caching minimum spanning trees as explained in
\autoref{sec:design/clustering}.

Since \name expects stack traces and \gls{asan} reports as input, we also
provide a supplemental script that runs the binary with a set of
\glspl{sci} (in parallel) and records the required data using GDB.
Together, our \name prototype and the data collection script allow developers
to use our deduplication approach directly on crashing inputs gathered from
fuzzing runs without any additional work.
Furthermore, thanks to \name's modular design, the prototype can be easily
adapted to further improve integration into real developing pipelines.
For instance, fuzzing tools might support the collection of the required
crash information (stack traces and \gls{asan} reports) during the fuzzing
phase without the need to reproduce the crash at a later time.
Moreover, since all steps except the final clustering can
be done independently for each crashing input, it might make sense to split
the \name workflow into two parts:
The first performs the preprocessing, computes the \gls{llm} embeddings and
the final vectors, and is hooked up directly to the fuzzer output so that
each crashing input is prepared right away.
The second part waits until all crashing inputs have been processed and
then computes the final clustering.

\section{Evaluation}
\label{sec:evaluation}

In this section, we aim to answer the following questions:
\begin{itemize}
    \item How well does \name perform at crash deduplication?
    \item How does the choice of data sources impact the quality of
          deduplication results?
    \item Which clustering algorithms are suitable for our approach?
    \item How does the choice of \gls{llm} affect deduplication quality?
    \item How does \name compare to existing state-of-the-art deduplication
          approaches?
    \item Do the underlying bug types influence the quality of deduplication
          results of \name and other approaches?
\end{itemize}
We evaluated our prototype implementation on the ground truth \gls{sci}
benchmark\footnote{
    \href{https://github.com/HexHive/Igor?tab=readme-ov-file\#ground-truth-benchmark}
    {https://github.com/HexHive/Igor?tab=readme-ov-file\#ground-truth-benchmark}
} created by the Igor~\cite{igor} authors.
It is based on the Magma~\cite{magma} and MoonLight~\cite{moonlight} benchmarks
and contains \num{344502} \glspl{sci} for 14 targets written in \cpp with
overall 61 ground truth labels grouping \glspl{sci} that are caused by the same
bug.
However, during their work on Igor, \citeauthor{igor} found that some labels
share the same root cause.
Moreover, we noticed that one label (tiffcp: CVE201610269) only contains
duplicate inputs that are also present under other labels
(tiffcp: AAH013 and tiffcp: AAH014, both caused by CVE-2016-10269).
We removed this ambiguous label and merged the duplicates identified by
the Igor authors, yielding \num{341952} inputs for 14 targets
associated with 54 distinct ground truth labels.

In the benchmark, \glspl{sci} are grouped in directories named according to the
corresponding ground truth label.
Since the filename of an \gls{sci} is sometimes contained in the resulting
stack traces, we flattened the directory structure and renamed all \gls{sci}
files using generic file names to prevent leakage of ground truth information
to the \gls{llm}, potentially (positively) disturbing the results.

During execution, some \glspl{sci} failed to reproduce
the claimed crash.
Out of the 54 distinct ground truth labels, we were not able to reproduce
crashes for four of them (tiff2pdf: D, tiffcp: AAH015, xmllint: C and
x509: AAH055).
In the end, we collected a total of \num{327071} stack traces and
\num{325044} \gls{asan} reports for 14 targets associated with 50 ground truth
labels.

The experiments were carried out on a machine running Debian 12 (Linux 6.9) with
two Intel Xeon Silver 4510 @ \qty{4.10}{\GHz} each with 12 cores,
\qty{512}{\giga\byte} of memory and an NVIDIA H100 NVL.
Unless explicitly stated otherwise, we used \tembed as \gls{llm}.

\subsection{Evaluation Metrics}

To get a better understanding of deduplication results, we use the
ground truth labeling for each target to compute \emph{purity},
\emph{inverse purity} and \emph{F-measure} as defined in \cite{purity}.
We briefly recall the definitions of these measures in the context of our
evaluation.
Consider a fixed target program containing $n$ bugs with ground truth labels
$l_1, \dots, l_n$ and suppose that $N$ \glspl{sci} were supplied to \name.
Let
\begin{displaymath}
    L_i \coloneqq
    \left\{\text{data points that belong to ground truth label } l_i\right\}
\end{displaymath}
for $i=1, \dots, n$.
Suppose further that $c_1, \dots, c_m$ are the generated cluster labels and
let
\begin{displaymath}
    C_j \coloneqq \left\{\text{data points that belong to cluster } c_j\right\},
\end{displaymath}
for $j=1, \dots, m$.
We define the precision
\begin{displaymath}
    \operatorname{Precision}(L_i, C_j) \coloneqq
    \frac{\abs*{L_i \cap C_j}}{\abs*{C_j}}
\end{displaymath}
and recall
\begin{displaymath}
    \operatorname{Recall}(L_i, C_j) \coloneqq
    \operatorname{Precision}(C_j, L_i) = \frac{\abs*{L_i \cap C_j}}{\abs*{L_i}}
\end{displaymath}
for the clustering $C_1, \dots, C_m$.
The purity
\begin{displaymath}
    \operatorname{Purity} \coloneqq \sum_{j=1}^m \frac{\abs*{C_j}}{N}
    \max_{i=1,\dots,n} \operatorname{Precision}(L_i, C_j)
\end{displaymath}
indicates whether data points in a cluster can be trusted to have the
same underlying bug or whether data points with different underlying bugs
are mixed into common clusters.
The inverse purity
\begin{displaymath}
    \operatorname{InversePurity} \coloneqq \sum_{i=1}^n \frac{\abs*{L_i}}{N}
    \max_{j=1,\dots,m} \operatorname{Recall}(L_i, C_j)
\end{displaymath}
on the other hand is a measure of how much data points with the same label
are spread out across multiple clusters.

A method that always creates a new cluster for each \gls{sci} will achieve
perfect purity scores of \qty{100}{\%}, whereas a method that consistently
groups all \glspl{sci} into a single cluster will have perfect inverse purity
scores of \qty{100}{\%}.
Therefore, one always needs to consider both purity and inverse purity.
The F-measure
\begin{displaymath}
    \operatorname{F-Measure} \coloneqq \sum_{i=1}^n \frac{\abs*{L_i}}{N}
    \max_{j=1,\dots,m} \operatorname{F}(L_i, C_j)
\end{displaymath}
combines purity and inverse purity into a single metric for quantifying the
quality of a deduplication method.
Here,
\begin{displaymath}
    \operatorname{F}(L_i, C_j) \coloneqq
    \frac{
        2 \operatorname{Precision}(L_i, C_j) \operatorname{Recall}(L_i, C_j)
    }{
        \operatorname{Precision}(L_i, C_j) + \operatorname{Recall}(L_i, C_j)
    },
\end{displaymath}
for $i=1, \dots, n$, $j=1, \dots, m$, where we set
$\operatorname{F}(L_i, C_j) \coloneqq 0$ in case both
$\operatorname{Precision}(L_i, C_j)$ and $\operatorname{Recall}(L_i, C_j)$
vanish.

\subsection{Results of \name}
\label{sec:evaluation/standard}

We present the results of \name per target in the evaluation dataset in the
\emph{All Data Sources} column of \autoref{tab:eval-data-sources}.
They show that \name predicts cluster counts close to the actual ground truth.
We observe high and often perfect purity values, indicating that \name
rarely groups together \glspl{sci} with different underlying bugs.
In practice, this allows developers to pick just one \gls{sci} from
each cluster for further analysis without missing any of the bugs.
High inverse purity scores suggest that in those cases where \name produces too
many clusters, those additional clusters mainly consist of a relatively small
number of noisy \glspl{sci} while the vast majority of \glspl{sci} for each
label is still concentrated in one main cluster.
Thus, by analyzing single \glspl{sci} from the larger clusters, a developer is
likely to cover all underlying bugs before reaching the smaller clusters that
contain duplicate \glspl{sci}.
Overall, \name produces clusterings that very accurately resemble the
ground truth labeling.
\begin{table*}[ht]
    \centering
    \caption{
        \name evaluation results for different choices of data sources using
        OpenAI's \tembed.
        Configurations:
        \emph{All Data Sources} is our proposed approach for \name combining
        full and coarse stack traces as well as \gls{asan} reports (see
        \autoref{sec:design}).
        \emph{No Full Traces} uses coarse traces and \gls{asan} reports but no
        full traces.
        \emph{No Coarse Traces} uses full traces and \gls{asan} reports but no
        coarse traces.
        \emph{No \gls{asan} Reports} uses just the full and coarse traces but no
        \gls{asan} reports.
        \emph{Only \gls{asan} Reports} uses just the \gls{asan} reports where
        the contained stack traces were not removed.
        \emph{Bugs} gives the number of ground truth bugs.
        The \emph{C} columns contain the number of clusters produced by the
        respective configurations.
        \emph{P}, \emph{IP}, and \emph{F} denote purity, inverse purity, and
        F-measure, given in percentages rounded to the nearest integer.
        Best entries per target and metric are highlighted in green.
    }
    \label{tab:eval-data-sources}
    \setlength\tabcolsep{3pt}
    \rowcolors{3}{lightgray}{}
    \newlength\customtabcolsep
    \setlength\customtabcolsep{6pt}
    \begin{tabular*}{\textwidth}{ l r@{\extracolsep{\fill}} r@{\extracolsep{\customtabcolsep}}r@{\extracolsep{\customtabcolsep}}r@{\extracolsep{\customtabcolsep}}r@{\extracolsep{\fill}} r@{\extracolsep{\customtabcolsep}}r@{\extracolsep{\customtabcolsep}}r@{\extracolsep{\customtabcolsep}}r@{\extracolsep{\fill}} r@{\extracolsep{\customtabcolsep}}r@{\extracolsep{\customtabcolsep}}r@{\extracolsep{\customtabcolsep}}r@{\extracolsep{\fill}} r@{\extracolsep{\customtabcolsep}}r@{\extracolsep{\customtabcolsep}}r@{\extracolsep{\customtabcolsep}}r@{\extracolsep{\fill}} r@{\extracolsep{\customtabcolsep}}r@{\extracolsep{\customtabcolsep}}r@{\extracolsep{\customtabcolsep}}r }
        \toprule
        \multirow{2}{*}{\thead{Target}}                     & \multirow{2}{*}{\thead{Bugs}}  & \multicolumn{4}{c}{\thead{All Data Sources}}                               & \multicolumn{4}{c}{\thead{No Full Traces}}                                 & \multicolumn{4}{c}{\thead{No Coarse Traces}}                               & \multicolumn{4}{c}{\thead{No \gls{asan} Reports}}                          & \multicolumn{4}{c}{\thead{Only \gls{asan} Reports}}
        \\                                                  &                                & \makecell[c]{C}  & \makecell[c]{P}  & \makecell[c]{IP}  & \makecell[c]{F}  & \makecell[c]{C}  & \makecell[c]{P}  & \makecell[c]{IP}  & \makecell[c]{F}  & \makecell[c]{C}  & \makecell[c]{P}  & \makecell[c]{IP}  & \makecell[c]{F}  & \makecell[c]{C}  & \makecell[c]{P}  & \makecell[c]{IP}  & \makecell[c]{F}  & \makecell[c]{C}  & \makecell[c]{P}  & \makecell[c]{IP}  & \makecell[c]{F}  \\
        \midrule
        char2svg                                            & 6                              & 8                & \bestcell{100}   & \bestcell{98}     & \bestcell{99}    & 12               & \bestcell{100}   & 86                & 92               & 9                & \bestcell{100}   & \bestcell{98}     & \bestcell{99}    & 7                & 93               & \bestcell{98}     & 91               & 9                & \bestcell{100}   & \bestcell{98}     & \bestcell{99}    \\
        client                                              & 1                              & 1                & \bestcell{100}   & \bestcell{100}    & \bestcell{100}   & 1                & \bestcell{100}   & \bestcell{100}    & \bestcell{100}   & 1                & \bestcell{100}   & \bestcell{100}    & \bestcell{100}   & 1                & \bestcell{100}   & \bestcell{100}    & \bestcell{100}   & 5                & \bestcell{100}   & 95                & 98               \\
        exif                                                & 1                              & 1                & \bestcell{100}   & \bestcell{100}    & \bestcell{100}   & 1                & \bestcell{100}   & \bestcell{100}    & \bestcell{100}   & 2                & \bestcell{100}   & 99                & 99               & 1                & \bestcell{100}   & \bestcell{100}    & \bestcell{100}   & 1                & \bestcell{100}   & \bestcell{100}    & \bestcell{100}   \\
        \makecell[l]{libxml2\_xml\_read\_\\memory\_fuzzer}  & 1                              & 1                & \bestcell{100}   & \bestcell{100}    & \bestcell{100}   & 1                & \bestcell{100}   & \bestcell{100}    & \bestcell{100}   & 1                & \bestcell{100}   & \bestcell{100}    & \bestcell{100}   & 1                & \bestcell{100}   & \bestcell{100}    & \bestcell{100}   & 1                & \bestcell{100}   & \bestcell{100}    & \bestcell{100}   \\
        pdf\_fuzzer                                         & 2                              & 2                & \bestcell{100}   & \bestcell{100}    & \bestcell{100}   & 2                & \bestcell{100}   & \bestcell{100}    & \bestcell{100}   & 2                & \bestcell{100}   & \bestcell{100}    & \bestcell{100}   & 2                & \bestcell{100}   & \bestcell{100}    & \bestcell{100}   & 2                & \bestcell{100}   & \bestcell{100}    & \bestcell{100}   \\
        pdfimages                                           & 4                              & 5                & \bestcell{100}   & 96                & 98               & 5                & \bestcell{100}   & 96                & 98               & 4                & \bestcell{100}   & \bestcell{100}    & \bestcell{100}   & 4                & \bestcell{100}   & \bestcell{100}    & \bestcell{100}   & 5                & \bestcell{100}   & 96                & 98               \\
        pdftoppm                                            & 3                              & 4                & \bestcell{100}   & \bestcell{100}    & \bestcell{100}   & 1                & 81               & \bestcell{100}    & 76               & 4                & \bestcell{100}   & \bestcell{100}    & \bestcell{100}   & 4                & \bestcell{100}   & \bestcell{100}    & \bestcell{100}   & 4                & \bestcell{100}   & \bestcell{100}    & \bestcell{100}   \\
        pdftotext                                           & 2                              & 2                & \bestcell{100}   & \bestcell{100}    & \bestcell{100}   & 2                & \bestcell{100}   & \bestcell{100}    & \bestcell{100}   & 2                & \bestcell{100}   & \bestcell{100}    & \bestcell{100}   & 4                & \bestcell{100}   & 91                & 95               & 2                & \bestcell{100}   & \bestcell{100}    & \bestcell{100}   \\
        sox (mp3)                                           & 7                              & 9                & \bestcell{100}   & 96                & 98               & 9                & \bestcell{100}   & 97                & 98               & 9                & \bestcell{100}   & 96                & 98               & 10               & \bestcell{100}   & 96                & 98               & 9                & \bestcell{100}   & \bestcell{100}    & \bestcell{100}   \\
        sox (wav)                                           & 5                              & 8                & \bestcell{100}   & \bestcell{56}     & 71               & 8                & \bestcell{100}   & \bestcell{56}     & 71               & 8                & \bestcell{100}   & \bestcell{56}     & \bestcell{72}    & 9                & \bestcell{100}   & \bestcell{56}     & 71               & 9                & \bestcell{100}   & \bestcell{56}     & 71               \\
        tiff2pdf                                            & 3                              & 5                & \bestcell{100}   & 97                & 98               & 3                & 99               & \bestcell{100}    & 99               & 36               & \bestcell{100}   & 91                & 94               & 5                & \bestcell{100}   & 97                & 98               & 4                & \bestcell{100}   & \bestcell{100}    & \bestcell{100}   \\
        tiffcp                                              & 6                              & 6                & \bestcell{85}    & \bestcell{99}     & \bestcell{89}    & 6                & \bestcell{85}    & \bestcell{99}     & \bestcell{89}    & 6                & \bestcell{85}    & \bestcell{99}     & 88               & 7                & \bestcell{85}    & 97                & 88               & 6                & \bestcell{85}    & \bestcell{99}     & \bestcell{89}    \\
        x509                                                & 1                              & 1                & \bestcell{100}   & \bestcell{100}    & \bestcell{100}   & 1                & \bestcell{100}   & \bestcell{100}    & \bestcell{100}   & 2                & \bestcell{100}   & \bestcell{100}    & \bestcell{100}   & 1                & \bestcell{100}   & \bestcell{100}    & \bestcell{100}   & 1                & \bestcell{100}   & \bestcell{100}    & \bestcell{100}   \\
        xmllint                                             & 8                              & 22               & \bestcell{83}    & 74                & 68               & 29               & \bestcell{83}    & 72                & 67               & 1                & 78               & \bestcell{100}    & \bestcell{74}    & 1                & 79               & \bestcell{100}    & \bestcell{74}    & 1                & 79               & \bestcell{100}    & \bestcell{74}    \\
        \rowcolor{white}
        \midrule
        Average                                             &                                &                  & \bestcell{98}    & 94                & 94               &                  & 96               & 93                & 92               &                  & 97               & \bestcell{96}     & \bestcell{95}    &                  & 97               & 95                & 94               &                  & 97               & \bestcell{96}     & \bestcell{95}    \\
        \bottomrule
    \end{tabular*}
\end{table*}

\subsection{Effect of Choice of Data Sources}
\label{sec:evaluation/data-sources}

To assess how the selection of data sources affects deduplication
quality, we used our proposed approach of combining \emph{All Data Sources}
as a baseline and then removed one of the data sources.
The results are shown in the center three columns of
\autoref{tab:eval-data-sources}.
We also evaluated configurations that only use a single data source and found
that out of those, the best deduplication quality is achieved when using
\emph{Only \gls{asan} Reports} (without removing the contained stack traces).
The good results of this latter configuration indicate that relying solely on
one data source can be sufficient for achieving accurate deduplication for many
targets.
Based on these results, there is only limited room for improvement when
evaluating other \name configurations and we see, in fact, that all assessed
configurations can be used to deduplicate \glspl{sci} reliably.
This indicates that the general approach of \name is very flexible.
We still propose combining \emph{All Data Sources} for the following reasons:
This configuration produces the highest purity values, indicating that
clusters rarely include \glspl{sci} with different underlying bugs, thus proving
the reliability of the bucketing for developers.
In contrast, leaving out full stack traces decreases all three metrics for
several targets.
This suggests that the information contained in the full stack traces is
valuable for deduplication and excludes the \emph{No Full Traces} configuration
as a viable alternative.
Although removing coarse stack traces achieves good average metrics,
it produces significantly worse results than our baseline for
\emph{tiff2pdf}, inaccurately producing the largest number of clusters in our
comparison, and \emph{xmllint}.
\pagebreak
For the latter target, while not optimal, our baseline produces a grouping that
is acceptable for further analysis.
Meanwhile, the \emph{No Coarse Traces} configuration generates a single cluster
that provides no information about the actual ground truth.
The same incorrect single-cluster results for \emph{xmllint} are also observed
for the \emph{No \gls{asan} Reports} and \emph{Only \gls{asan} Reports}
configurations and they artificially increase the average inverse purity of
these configurations.
In fact, \emph{xmllint} is the sole reason why these achieve slightly better
average inverse purity than our baseline configuration and excluding this target
from the average rectifies this statistical effect.
Overall, we see that, while the \name method is able to produce very accurate
bucketings using different selections of data sources, combining
\emph{All Data Sources} is the only configuration that achieves robust and
usable results for all targets in our evaluation dataset.

\subsection{Effect of Clustering Algorithm}
\label{sec:evaluation/clustering}

\enlargethispage{-0.475\baselineskip}
We evaluated the performance of \name when using different clustering
algorithms and present the results in the three leftmost result columns of
\autoref{tab:eval-clustering-model}.
We compared the \emph{Hybrid HDBSCAN}~\cite{hybrid_hdbscan} approach described
in \autoref{sec:design/clustering} with \emph{DBSCAN}~\cite{dbscan} and
\emph{k-means}~\cite{kmeans}.
As the DBCV~\cite{dbcv} score is designed specifically for density-based
clustering algorithms, we did not use it for $k$-means and instead relied on
silhouette scores~\cite{silhouette} in this case.
Moreover, as mentioned earlier, $k$-means requires the number of clusters
to be specified in advance and the correct value can reach from $1$ to the
number of \glspl{sci}.
As the latter is very large for most of the targets in our dataset, we
restrict the parameter search to a maximum of \num{100} clusters.
\begin{table*}[ht]
    \centering
    \caption{
        \name evaluation results for different clustering algorithms and
        embedding models.
        Clustering algorithms:
        \emph{Hybrid HDBSCAN}~\cite{hybrid_hdbscan} as explained in
        \autoref{sec:design/clustering}, \emph{DBSCAN}~\cite{dbscan} and
        \emph{k-means}~\cite{kmeans}.
        Embedding models:
        \emph{\tembed} from OpenAI~\cite{openai_embedding},
        \emph{\nvembed} from NVIDIA~\cite{nvidia_embedding} and
        \emph{\stella}~\cite{stella_embedding}.
        \emph{Bugs} gives the number of ground truth bugs.
        The \emph{C} columns contain the number of clusters produced by the
        respective configurations.
        \emph{P}, \emph{IP}, and \emph{F} denote purity, inverse purity, and
        F-measure, given in percentages rounded to the nearest integer.
        Best entries per target and metric are highlighted in green.
        In the overlapping middle column, an entry is highlighted in lighter
        green with a mark in the bottom right corner if the configuration is the
        best among the embedding models but not among the clustering algorithms.
    }
    \label{tab:eval-clustering-model}
    \setlength\tabcolsep{3pt}
    \rowcolors{3}{}{lightgray}
    \setlength\customtabcolsep{6pt}
    \begin{tabular*}{\textwidth}{ l@{} r@{\extracolsep{\fill}} r@{\extracolsep{\customtabcolsep}}r@{\extracolsep{\customtabcolsep}}r@{\extracolsep{\customtabcolsep}}r@{\extracolsep{\fill}} r@{\extracolsep{\customtabcolsep}}r@{\extracolsep{\customtabcolsep}}r@{\extracolsep{\customtabcolsep}}r@{\extracolsep{\fill}} r@{\extracolsep{\customtabcolsep}}r@{\extracolsep{\customtabcolsep}}r@{\extracolsep{\customtabcolsep}}r@{\extracolsep{\fill}} r@{\extracolsep{\customtabcolsep}}r@{\extracolsep{\customtabcolsep}}r@{\extracolsep{\customtabcolsep}}r r@{\extracolsep{\customtabcolsep}}r@{\extracolsep{\customtabcolsep}}r@{\extracolsep{\customtabcolsep}}r }
        \toprule
        \multirow{3}{*}{\thead{Target}}                     & \multicolumn{1}{c}{\multirow{3}{*}{\thead{Bugs}}} & \multicolumn{4}{c}{\thead{DBSCAN}}                                         & \multicolumn{4}{c}{\thead{$k$-means}}                                      & \multicolumn{4}{c}{\thead{Hybrid HDBSCAN}}                                      &                  &                   &                   &                  &                  &                  &                   &
        \\                                                  &                                                   &                  &                  &                   &                  &                  &                  &                   &                  & \multicolumn{4}{c}{\thead{\tembedshort}}                                        & \multicolumn{4}{c}{\thead{\nvembed}}                                        & \multicolumn{4}{c}{\thead{\stella}}
        \\                                                  &                                                   & \makecell[c]{C}  & \makecell[c]{P}  & \makecell[c]{IP}  & \makecell[c]{F}  & \makecell[c]{C}  & \makecell[c]{P}  & \makecell[c]{IP}  & \makecell[c]{F}  & \makecell[c]{C}  & \makecell[c]{P}  & \makecell[c]{IP}    & \makecell[c]{F}     & \makecell[c]{C}  & \makecell[c]{P}   & \makecell[c]{IP}  & \makecell[c]{F}  & \makecell[c]{C}  & \makecell[c]{P}  & \makecell[c]{IP}  & \makecell[c]{F}  \\
        \midrule
        char2svg                                            & 6                                                 & 8                & \bestcellv{100}  & 98                & \bestcellv{99}   & 2                & 72               & \bestcellv{100}   & 73               & 8                & \bestcellv{100}  & \bestcellright{98}  & \bestcellv{99}      & 8                &  \bestcellv{100}  & \bestcellv{98}    & \bestcellv{99}   & 7                & 93               & \bestcellv{98}    & 91               \\
        client                                              & 1                                                 & 1                & \bestcellv{100}  & \bestcellv{100}   & \bestcellv{100}  & 1                & \bestcellv{100}  & \bestcellv{100}   & \bestcellv{100}  & 1                & \bestcellv{100}  & \bestcellv{100}     & \bestcellv{100}     & 1                &  \bestcellv{100}  & \bestcellv{100}   & \bestcellv{100}  & 1                & \bestcellv{100}  & \bestcellv{100}   & \bestcellv{100}  \\
        exif                                                & 1                                                 & 1                & \bestcellv{100}  & \bestcellv{100}   & \bestcellv{100}  & 20               & \bestcellv{100}  & 34                & 51               & 1                & \bestcellv{100}  & \bestcellv{100}     & \bestcellv{100}     & 1                &  \bestcellv{100}  & \bestcellv{100}   & \bestcellv{100}  & 1                & \bestcellv{100}  & \bestcellv{100}   & \bestcellv{100}  \\
        \makecell[l]{libxml2\_xml\_read\_\\memory\_fuzzer}  & 1                                                 & 1                & \bestcellv{100}  & \bestcellv{100}   & \bestcellv{100}  & 1                & \bestcellv{100}  & \bestcellv{100}   & \bestcellv{100}  & 1                & \bestcellv{100}  & \bestcellv{100}     & \bestcellv{100}     & 1                &  \bestcellv{100}  & \bestcellv{100}   & \bestcellv{100}  & 1                & \bestcellv{100}  & \bestcellv{100}   & \bestcellv{100}  \\
        pdf\_fuzzer                                         & 2                                                 & 2                & \bestcellv{100}  & \bestcellv{100}   & \bestcellv{100}  & 2                & \bestcellv{100}  & \bestcellv{100}   & \bestcellv{100}  & 2                & \bestcellv{100}  & \bestcellv{100}     & \bestcellv{100}     & 2                &  \bestcellv{100}  & \bestcellv{100}   & \bestcellv{100}  & 2                & \bestcellv{100}  & \bestcellv{100}   & \bestcellv{100}  \\
        pdfimages                                           & 4                                                 & 5                & \bestcellv{100}  & 96                & \bestcellv{98}   & 3                & 81               & \bestcellv{100}   & 87               & 5                & \bestcellv{100}  & \bestcellright{96}  & \bestcellv{98}      & 4                &  96               & \bestcellv{96}    & 96               & 4                & 96               & \bestcellv{96}    & 96               \\
        pdftoppm                                            & 3                                                 & 4                & \bestcellv{100}  & \bestcellv{100}   & \bestcellv{100}  & 3                & \bestcellv{100}  & \bestcellv{100}   & \bestcellv{100}  & 4                & \bestcellv{100}  & \bestcellv{100}     & \bestcellv{100}     & 4                &  \bestcellv{100}  & \bestcellv{100}   & \bestcellv{100}  & 4                & \bestcellv{100}  & \bestcellv{100}   & \bestcellv{100}  \\
        pdftotext                                           & 2                                                 & 5                & \bestcellv{100}  & 87                & 92               & 2                & \bestcellv{100}  & \bestcellv{100}   & \bestcellv{100}  & 2                & \bestcellv{100}  & \bestcellv{100}     & \bestcellv{100}     & 4                &  \bestcellv{100}  & 91                & 95               & 1                & 85               & \bestcellv{100}   & 82               \\
        sox (mp3)                                           & 7                                                 & 10               & \bestcellv{100}  & \bestcellv{96}    & \bestcellv{98}   & 9                & \bestcellv{100}  & \bestcellv{96}    & \bestcellv{98}   & 9                & \bestcellv{100}  & \bestcellv{96}      & \bestcellv{98}      & 10               &  \bestcellv{100}  & \bestcellv{96}    & \bestcellv{98}   & 10               & \bestcellv{100}  & \bestcellv{96}    & \bestcellv{98}   \\
        sox (wav)                                           & 5                                                 & 10               & \bestcellv{100}  & 56                & 71               & 6                & \bestcellv{100}  & \bestcellv{99}    & \bestcellv{100}  & 8                & \bestcellv{100}  & \bestcellright{56}  & \bestcellright{71}  & 8                &  \bestcellv{100}  & \bestcellv{56}    & \bestcellv{71}   & 8                & \bestcellv{100}  & \bestcellv{56}    & \bestcellv{71}   \\
        tiff2pdf                                            & 3                                                 & 6                & \bestcellv{100}  & 97                & 98               & 3                & 99               & \bestcellv{100}   & \bestcellv{99}   & 5                & \bestcellv{100}  & 97                  & 98                  & 5                &  \bestcellv{100}  & \bestcellv{100}   & \bestcellv{100}  & 1                & 93               & \bestcellv{100}   & 90               \\
        tiffcp                                              & 6                                                 & 6                & \bestcellv{85}   & 99                & \bestcellv{89}   & 4                & 67               & \bestcellv{100}   & 70               & 6                & \bestcellv{85}   & \bestcellright{99}  & \bestcellv{89}      & 7                &  \bestcellv{85}   & 97                & 88               & 9                & \bestcellv{85}   & 87                & 85               \\
        x509                                                & 1                                                 & 1                & \bestcellv{100}  & \bestcellv{100}   & \bestcellv{100}  & 1                & \bestcellv{100}  & \bestcellv{100}   & \bestcellv{100}  & 1                & \bestcellv{100}  & \bestcellv{100}     & \bestcellv{100}     & 1                &  \bestcellv{100}  & \bestcellv{100}   & \bestcellv{100}  & 1                & \bestcellv{100}  & \bestcellv{100}   & \bestcellv{100}  \\
        xmllint                                             & 8                                                 & 1                & 79               & \bestcellv{100}   & \bestcellv{74}   & 6                & \bestcellv{83}   & 77                & 69               & 22               & \bestcellv{83}   & 74                  & 68                  & 1                &  79               & \bestcellv{100}   & \bestcellv{74}   & 1                & 79               & \bestcellv{100}   & \bestcellv{74}   \\
        \rowcolor{white}
        \midrule
        Average                                             &                                                   &                  & 97               & \bestcellv{95}    & \bestcellv{94}   &                  & 93               & 93                & 89               &                  & \bestcellv{98}   & 94                  & \bestcellv{94}      &                  &  97               & \bestcellv{95}    & \bestcellv{94}   &                  & 95               & \bestcellv{95}    & 92               \\
        \bottomrule
    \end{tabular*}%
\end{table*}

Both DBSCAN and HDBSCAN produce good results for most targets, but
overall the results favor HDBSCAN, which is more accurate for several targets
(e.g., \emph{pdftotext} and \emph{tiff2pdf}).
While $k$-means also yields acceptable deduplication quality, the average metrics
indicate a noticeably worse performance than the density-based clustering
approaches.
Depending on the target, we observe problems regarding both purity
(\eg \emph{char2svg} and \emph{tiffcp}) and inverse purity (\eg \emph{exif})
of the produced bucketings.
Combined with the practical difficulty of searching for the best number of
clusters, this suggests that, as suspected, $k$-means is not suitable for our
use case.

\subsection{Effect of Choice of Embedding Model}
\label{sec:evaluation/model}

In addition to using OpenAI's \tembed model, we also evaluated
\glspl{llm} that can be run locally.
We chose NVIDIA's \nvembed~\cite{nvidia_embedding} because it is among the
highest-ranking models on the \gls{mteb}
leaderboard~\cite{embedding_leaderboard} (as of January 13, 2025), which ranks
the performance of embedding models across different tasks.
We also tried \stella~\cite{stella_embedding} because it ranks high on the
\gls{mteb} leaderboard as well while being substantially smaller
(\textasciitilde\qty{1.5}{\billion} parameters) than NVIDIA's \nvembed
(\textasciitilde\qty{8}{\billion} parameters) and thus suitable for less
powerful hardware.

\begin{figure}[b]
    \centering
    \begin{subfigure}{0.45\columnwidth}
        \includegraphics[width=\textwidth]{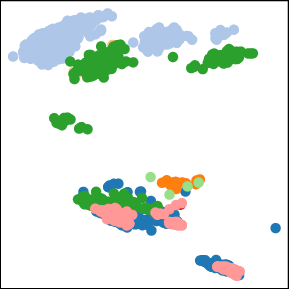}
        \caption*{\tembed}
    \end{subfigure}
    \hfill
    \begin{subfigure}{0.45\columnwidth}
        \includegraphics[width=\textwidth]{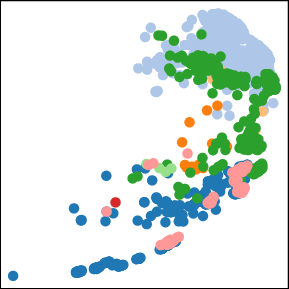}
        \caption*{\nvembed}
    \end{subfigure}
    \caption{
        Two-dimensional projections of the combined vectors for xmllint,
        obtained using scikit-learn's~\cite{sklearn,sklearn_python} truncated
        singular value decomposition.
        The colors indicate the eight different ground truth labels.
    }
    \label{fig:model-comparison}
    \Description{
        Two-dimensional projections of the combined vectors for xmllint, once
        when using \tembed and once using \nvembed.
        The \tembed projection shows a clustering that is more clearly in line
        with the ground truth labeling while the \nvembed projection shows a
        more chaotic distribution.
    }
\end{figure}

The three rightmost columns of \autoref{tab:eval-clustering-model} show that for
most targets all three models produce good estimates of the ground truth,
indicating that \name can be used with different \glspl{llm}.
However, we see some decline in deduplication quality when using
the smaller \stella model suggesting that \name can profit from larger
models like \tembed or \nvembed.

The target where the difference between the evaluated models is most noticeable
is xmllint.
While the results generated using \tembed remain reasonably accurate, both
\nvembed and \stella produce poor results.
The \glspl{sci} for xmllint generate long execution paths,
many of which differ only marginally even across ground truth labels.
The results in \autoref{sec:evaluation/comparison} suggest that this also poses
a problem for other deduplication approaches.
While \tembed still extracts distinguishing information from the coarse stack
traces, this information does not seem pronounced enough in the embeddings
from \nvembed and \stella.
This limitation persists in the combined embeddings, as can be seen in the
two-dimensional projections in \autoref{fig:model-comparison}.
Consequently, \name produces useful clusters with \tembed but fails to
do so with the other models.

The choice of \gls{llm} also impacts the run time of \name.
Using OpenAI's batch API, results of embedding requests are claimed to be
returned within 24 hours.
In practice, we have experienced significantly shorter turnaround times, ranging
from under a minute to a few hours, depending on the number of embeddings
requested.
Still, gathering embeddings constitutes the majority of \name's run time.
This is evidently influenced by OpenAI's capacity at the time of the request and
while we did not perform a systematic evaluation, we observed that deduplicating
the entire dataset corresponding to \num{327071} \glspl{sci} using \tembed
is possible in less than ten hours.
In contrast, utilizing a local \gls{llm} on suitable hardware significantly
reduces run times.
On our evaluation machine, deduplication with \stella takes only
41 minutes, while using \nvembed reduces the run time to 39 minutes.

\subsection{Comparison to Existing Deduplication Methods}
\label{sec:evaluation/comparison}
\begin{table*}[ht]
    \centering
    \caption{
        Deduplication results of \emph{\name} in comparison to
        \emph{Crashwalk}~\cite{crashwalk}, our implementation of
        \emph{DeFault}~\cite{default}, and \emph{Igor}~\cite{igor}.
        \emph{Bugs} gives the number of ground truth bugs.
        The \emph{C} columns contain the number of clusters produced by the
        respective configurations.
        \emph{P}, \emph{IP}, and \emph{F} denote purity, inverse purity, and
        F-measure, given in percentages rounded to the nearest integer.
        Best entries per target and metric are highlighted in green.
    }
    \label{tab:eval-comparison}
    \setlength\tabcolsep{5.3pt}
    \setlength\customtabcolsep{10.6pt}
    \rowcolors{3}{lightgray}{}
    \begin{tabular*}{\textwidth}{ l r@{\extracolsep{\fill}} r@{\extracolsep{\customtabcolsep}}r@{\extracolsep{\customtabcolsep}}r@{\extracolsep{\customtabcolsep}}r@{\extracolsep{\fill}} r@{\extracolsep{\customtabcolsep}}r@{\extracolsep{\customtabcolsep}}r@{\extracolsep{\customtabcolsep}}r@{\extracolsep{\fill}} r@{\extracolsep{\customtabcolsep}}r@{\extracolsep{\customtabcolsep}}r@{\extracolsep{\customtabcolsep}}r@{\extracolsep{\fill}} r@{\extracolsep{\customtabcolsep}}r@{\extracolsep{\customtabcolsep}}r@{\extracolsep{\customtabcolsep}}r }
        \toprule
        \multirow{2}{*}{\thead{Target}}                     & \multirow{2}{*}{\thead{Bugs}}  & \multicolumn{4}{c}{\thead{\name}}                                           & \multicolumn{4}{c}{\thead{Crashwalk}}                                      & \multicolumn{4}{c}{\thead{DeFault}}                                        & \multicolumn{4}{c}{\thead{Igor}}
        \\                                                  &                                & \makecell[c]{C}   & \makecell[c]{P}  & \makecell[c]{IP}  & \makecell[c]{F}  & \makecell[c]{C}  & \makecell[c]{P}  & \makecell[c]{IP}  & \makecell[c]{F}  & \makecell[c]{C}  & \makecell[c]{P}  & \makecell[c]{IP}  & \makecell[c]{F}  & \makecell[c]{C}  & \makecell[c]{P}  & \makecell[c]{IP}  & \makecell[c]{F}  \\
        \midrule
        char2svg                                            & 6                              & 8                 & \bestcell{100}   & 98                & \bestcell{99}    & 67               & \bestcell{100}   & 61                & 73               & 2                & 37               & \bestcell{99}     & 46               & 3                & 33               & 91                & 34               \\
        client                                              & 1                              & 1                 & \bestcell{100}   & \bestcell{100}    & \bestcell{100}   & 1                & \bestcell{100}   & \bestcell{100}    & \bestcell{100}   & 1                & \bestcell{100}   & \bestcell{100}    & \bestcell{100}   & 1                & \bestcell{100}   & \bestcell{100}    & \bestcell{100}   \\
        exif                                                & 1                              & 1                 & \bestcell{100}   & \bestcell{100}    & \bestcell{100}   & 121              & \bestcell{100}   & 32                & 49               & 2                & \bestcell{100}   & \bestcell{100}    & \bestcell{100}   & 2                & \bestcell{100}   & 61                & 76               \\
        \makecell[l]{libxml2\_xml\_read\_\\memory\_fuzzer}  & 1                              & 1                 & \bestcell{100}   & \bestcell{100}    & \bestcell{100}   & 99               & \bestcell{100}   & 11                & 20               & 1                & \bestcell{100}   & \bestcell{100}    & \bestcell{100}   & 3                & \bestcell{100}   & 38                & 55               \\
        pdf\_fuzzer                                         & 2                              & 2                 & \bestcell{100}   & \bestcell{100}    & \bestcell{100}   & 10               & \bestcell{100}   & 54                & 68               & 1                & 72               & \bestcell{100}    & 73               & 2                & \bestcell{100}   & \bestcell{100}    & \bestcell{100}   \\
        pdfimages                                           & 4                              & 5                 & \bestcell{100}   & \bestcell{96}     & \bestcell{98}    & 21               & \bestcell{100}   & 86                & 90               & 5                & 46               & 87                & 45               & 6                & 87               & 72                & 78               \\
        pdftoppm                                            & 3                              & 4                 & \bestcell{100}   & \bestcell{100}    & \bestcell{100}   & 6                & \bestcell{100}   & \bestcell{100}    & \bestcell{100}   & 4                & 82               & 83                & 78               & 3                & 65               & 81                & 67               \\
        pdftotext                                           & 2                              & 2                 & \bestcell{100}   & \bestcell{100}    & \bestcell{100}   & 10               & \bestcell{100}   & 88                & 91               & 2                & 93               & 97                & 91               & 3                & 90               & 56                & 62               \\
        sox (mp3)                                           & 7                              & 9                 & \bestcell{100}   & 96                & \bestcell{98}    & 18               & \bestcell{100}   & 96                & \bestcell{98}    & 2                & 92               & 97                & 87               & 2                & 42               & \bestcell{99}     & 51               \\
        sox (wav)                                           & 5                              & 8                 & \bestcell{100}   & 56                & 71               & 15               & \bestcell{100}   & 55                & 71               & 1                & 95               & \bestcell{100}    & \bestcell{93}    & 2                & 60               & \bestcell{100}    & 68               \\
        tiff2pdf                                            & 3                              & 5                 & \bestcell{100}   & 97                & \bestcell{98}    & 7                & \bestcell{100}   & 89                & 94               & 1                & 93               & \bestcell{100}    & 91               & 2                & 49               & 72                & 56               \\
        tiffcp                                              & 6                              & 6                 & 85               & \bestcell{99}     & \bestcell{89}    & 12               & 85               & 74                & 76               & 4                & 59               & 91                & 63               & 6                & \bestcell{89}    & 90                & \bestcell{89}    \\
        x509                                                & 1                              & 1                 & \bestcell{100}   & \bestcell{100}    & \bestcell{100}   & 1                & \bestcell{100}   & \bestcell{100}    & \bestcell{100}   & 1                & \bestcell{100}   & \bestcell{100}    & \bestcell{100}   & 1                & \bestcell{100}   & \bestcell{100}    & \bestcell{100}   \\
        xmllint                                             & 8                              & 22                & 83               & 74                & 68               & 825              & \bestcell{84}    & 18                & 28               & 1                & 79               & \bestcell{100}    & 75               & 4                & 70               & 98                & \bestcell{76}    \\
        \rowcolor{white}
        \midrule
        Average                                             &                                &                   & \bestcell{98}    & 94                & \bestcell{94}    &                  & \bestcell{98}    & 69                & 76               &                  & 82               & \bestcell{97}     & 82               &                  & 78               & 83                & 72               \\
        \bottomrule
    \end{tabular*}
\end{table*}
Having determined the best-performing configuration for \name, we now compare
its performance to that of Crashwalk~\cite{crashwalk}, which serves
as a representative of stack hashing approaches, DeFault~\cite{default}, which
groups crashes based on mutual information scores of the basic blocks
present in their execution traces, and Igor~\cite{igor}, which compares
minimized control flow graphs of crashes.
We do this by, again, collecting the number of clusters as well as purity,
inverse purity and F-measure.
The results are presented in \autoref{tab:eval-comparison}.

Crashwalk records stack traces of \glspl{sci} and groups together those inputs
whose stack traces have the same hash value.
We see from the results in \autoref{tab:eval-comparison} that, compared to
Crashwalk, \name provides more accurate deduplication results across almost all
targets.
Since crashes that result from different bugs are unlikely to produce identical
stack traces, Crashwalk is able to distinguish between different bugs very well,
as is expressed by its high purity scores.
However, even small differences in the stack traces lead to different hash
values and Crashwalk has no way of capturing such nuances.
This results in a lot of duplicate buckets for the same bug and hence lower
inverse purity scores.
\name, on the other hand, can take account of the similarity of both stack
traces as well as further information from the supplied \gls{asan} reports.
This leads to considerably higher inverse purity scores by virtue of lower
numbers of clusters that are much closer to the ground truth.

DeFault requires as input the basic block-level execution traces of the
crashing \glspl{sci} and of non-crashing inputs.
It then identifies basic blocks that have high mutual information scores related
to crashing behavior and groups the \glspl{sci} based on which basic blocks
their traces contain.
As the source code for DeFault is not publicly available, we implemented a
prototype ourselves based on the paper's description, focusing solely on the
deduplication mechanism while excluding unrelated components, such as fault
localization.\footnote{
    Our prototype implementation of DeFault is contained in our artifact
    repository at \repourl.
}
We then generated non-crashing inputs using AFL/\aflpp~\cite{afl, aflplusplus}
and collected basic block traces using Intel Pin~\cite{intel-pin}.
The results show that DeFault is not able to provide deduplication results that
are as accurate as those of \name.
While inverse purity scores are high, this is, again, often due to the small
number of clusters produced by DeFault, typically less than the number of
ground truth labels.
This also leads to groups of \glspl{sci} with different underlying
bugs, resulting in low purity scores.

Igor is a deduplication method that collects minimized control flow graphs of
\glspl{sci} and clusters them based on graph similarity metrics.
We evaluated Igor using the available prototype implementation and used a
running time of 15 minutes per \gls{sci} for the minimum-coverage fuzzing
phase as recommended by the authors.
As Igor's clustering algorithm is only capable of producing results with
at least two clusters, the authors suggest comparing their results to those
of a stack trace hashing approach, and choosing the result that produces the
smaller number of clusters.
We followed this recommendation leading to improved results for \emph{client}
and \emph{x509}.
For the other two single-bug targets, however, stack trace hashing generates
far too many clusters, making this strategy ineffective.
Even though Igor's results in these cases are still acceptable, they do not
recover the ground truth as reliably as \name.
Igor's deduplication also falls short for many of the other targets,
suffering from both low purity and inverse purity scores.
In general, Igor's results are not as accurate as those of \name.

\begin{table*}[!ht]
    \centering
    \caption{
        Results of \emph{\name}, \emph{Crashwalk}~\cite{crashwalk},
        our implementation of \emph{DeFault}~\cite{default}, and
        \emph{Igor}~\cite{igor} for different bug types classified by their
        \emph{CWE Type}~\cite{cwe}.
        The \emph{Total} column contains the number of bugs of each type.
        \emph{Over} columns contain the mean (\emph{$\mu$}) and standard
        deviation (\emph{$\sigma$}) of the overcounting scores for the
        respective bug types while \emph{Under} columns contain
        the mean and standard deviation of the undercounting scores,
        all rounded to one decimal place.
        Apparent outliers are highlighted in red.
    }
    \label{tab:eval-bug-types}
    \setlength\tabcolsep{3.8pt}
    \setlength\customtabcolsep{7.6pt}
    \begin{tabular*}{\textwidth}{ p{15em} r@{\extracolsep{\fill}} r@{\extracolsep{\customtabcolsep}}r@{\extracolsep{\customtabcolsep}}r@{\extracolsep{\customtabcolsep}}r@{\extracolsep{\fill}} r@{\extracolsep{\customtabcolsep}}r@{\extracolsep{\customtabcolsep}}r@{\extracolsep{\customtabcolsep}}r@{\extracolsep{\fill}} r@{\extracolsep{\customtabcolsep}}r@{\extracolsep{\customtabcolsep}}r@{\extracolsep{\customtabcolsep}}r@{\extracolsep{\fill}} r@{\extracolsep{\customtabcolsep}}r@{\extracolsep{\customtabcolsep}}r@{\extracolsep{\customtabcolsep}}r }
        \toprule
        \multirow{3}{*}{CWE Type}                                                & \multirow{3}{*}{\thead{Total}}  & \multicolumn{4}{c}{\thead{\name}}                                                               & \multicolumn{4}{c}{\thead{Crashwalk}}                                                           & \multicolumn{4}{c}{\thead{DeFault}}                                                              & \multicolumn{4}{c}{\thead{Igor}}
        \\                                                                       &                                 & \multicolumn{2}{c}{Over}                       & \multicolumn{2}{c}{Under}                      & \multicolumn{2}{c}{Over}                       & \multicolumn{2}{c}{Under}                      & \multicolumn{2}{c}{Over}                       & \multicolumn{2}{c}{Under}                       & \multicolumn{2}{c}{Over}                        & \multicolumn{2}{c}{Under}
        \\                                                                       &                                 & \makecell[c]{$\mu$}  & \makecell[c]{$\sigma$}  & \makecell[c]{$\mu$}  & \makecell[c]{$\sigma$}  & \makecell[c]{$\mu$}  & \makecell[c]{$\sigma$}  & \makecell[c]{$\mu$}  & \makecell[c]{$\sigma$}  & \makecell[c]{$\mu$}  & \makecell[c]{$\sigma$}  & \makecell[c]{$\mu$}  & \makecell[c]{$\sigma$}   & \makecell[c]{$\mu$}  & \makecell[c]{$\sigma$}   & \makecell[c]{$\mu$}  & \makecell[c]{$\sigma$}  \\ \rowcolor{lightgray}
        \midrule
        Improper Restriction of Operations within the Bounds of a Memory Buffer  & 28                              & 1.1                  & 2.7                     & 0.6                  & 1.0                     & 46.3                 & 89.1                    & 0.5                  & 0.7                     & 0.5                  & 0.8                     & 4.4                  & 2.2                      & 0.8                  & 0.9                      & 2.8                  & 1.7                     \\ \rowcolor{lightgray}
        \hspace{1em}Out-of-bounds Read                                           & 11                              & 0.3                  & 0.7                     & 0.2                  & 0.4                     & \outliercell{15.2}   & \outliercell{35.7}      & 0.2                  & 0.4                     & 1.1                  & 0.9                     & 3.3                  & 1.7                      & 1.2                  & 1.0                      & 2.5                  & 1.6                     \\ \rowcolor{lightgray}
        \hspace{1em}Out-of-bounds Write                                          & 6                               & 0.3                  & 0.5                     & 0.8                  & 1.0                     & 1.7                  & 0.8                     & 0.8                  & 1.0                     & 0.3                  & 0.8                     & 4.8                  & 1.0                      & 0.2                  & 0.4                      & 3.3                  & 1.4                     \\ \rowcolor{lightgray}
        \hspace{1em}Other                                                        & 11                              & \outliercell{2.3}    & \outliercell{4.0}       & 1                    & 1.3                     & \outliercell{101.6}  & \outliercell{120.5}     & 0.6                  & 0.7                     & 0                    & 0.0                     & 5.3                  & 2.7                      & 0.6                  & 0.8                      & 2.8                  & 2.0                     \\ \rowcolor{white}
        Incorrect Calculation                                                    & 11                              & 0.6                  & 0.9                     & 0.6                  & 0.8                     & 14.7                 & 35.2                    & 0.6                  & 0.8                     & 0.5                  & 0.9                     & 3.4                  & 2.2                      & 0.5                  & 0.7                      & 2                    & 1.7                     \\ \rowcolor{white}
        \hspace{1em}Integer Overflow or Wraparound                               & 10                              & 0.6                  & 1.0                     & 0.7                  & 0.8                     & \outliercell{16.1}   & \outliercell{36.8}      & 0.6                  & 0.8                     & 0.5                  & 1.0                     & 3.3                  & 2.3                      & 0.5                  & 0.7                      & 2.1                  & 1.8                     \\ \rowcolor{white}
        \hspace{1em}Divide By Zero                                               & 1                               & 1                    & 0                       & 0                    & 0                       & 1                    & 0                       & 0                    & 0                       & 0                    & 0                       & 4                    & 0                        & 0                    & 0                        & 1                    & 0                       \\ \rowcolor{lightgray}
        NULL Pointer Dereference                                                 & 6                               & 1                    & 1.6                     & 0.7                  & 0.8                     & 2.7                  & 2.3                     & 0.5                  & 0.8                     & 0.3                  & 0.5                     & 4.3                  & 2.1                      & 0.2                  & 0.4                      & 2.7                  & 2.1                     \\ \rowcolor{white}
        Other                                                                    & 15                              & 0.5                  & 1.1                     & 0.7                  & 1.4                     & 1.5                  & 1.5                     & 0.5                  & 0.7                     & 0.8                  & 1.2                     & 3.8                  & 2.1                      & 0.6                  & 0.7                      & 2.3                  & 1.6                     \\ \rowcolor{white}
        \midrule
        Total                                                                    & 60                              & 0.8                  & 2.0                     & 0.7                  & 1.0                     & 24.9                 & 65.3                    & 0.5                  & 0.7                     & 0.6                  & 0.9                     & 4.1                  & 2.2                      & 0.6                  & 0.8                      & 2.5                  & 1.7                     \\
        \bottomrule
    \end{tabular*}
\end{table*}

Apart from improved deduplication quality, \name also has the advantage of
requiring less data that is easier to obtain, compared to Igor and DeFault.
While \name only needs stack traces and \gls{asan} reports, both of which can
easily be collected during the fuzzing process, the other approaches both
require the collection of full execution traces.
These cannot be generated as part of the fuzzing process, as this would add
considerable overhead.
Furthermore, these full execution traces are oftentimes very large (the Igor
authors report sizes of up to \qty{10}{\gibi\byte} for a single trace),
potentially leading to difficulties in storing them and causing long run times
of the deduplication tools.
\pagebreak
The problem of long run times is especially relevant for Igor, whose
aforementioned minimum-coverage fuzzing phase is recommended by the authors to
run for 15 minutes per \gls{sci}.
On our dataset, this phase alone would require close to 4000 CPU hours
and thus multiple days on our evaluation machine.

\subsection{Influence of Bug Type}

In order to determine if the types of the underlying bugs influence the
deduplication quality of \name and the other approaches, we gathered the
CWE~\cite{cwe} classifications of the bugs in our evaluation dataset, according
to their CVE entries~\cite{cve}, and collected how the \glspl{sci} corresponding
to each of the bugs are bucketed by the various deduplication methods.
We present the results in \autoref{tab:eval-bug-types} where we measure the
deduplication quality for individual bugs by over- and undercounting scores.
Given a ground truth label $l$, the overcounting score
\begin{displaymath}
    \operatorname{Overcounting}(l) \coloneqq \abs*{\left\{\text{clusters containing an \gls{sci} from } l\right\}} - 1
\end{displaymath}
measures how many superfluous clusters are created for the label $l$, while the
undercounting score
\begin{displaymath}
    \operatorname{Undercounting}(l) \coloneqq \abs*{\left\{\begin{array}{c}\text{labels } l' \neq l \text{ for which some \gls{sci}}\\\text{shares a cluster with an \gls{sci} from } l\end{array}\right\}}
\end{displaymath}
counts how many other labels are grouped together into clusters with $l$.
Scores close to zero indicate accurate deduplication.

\enlargethispage{-0.7\baselineskip}
The results show, again, that Crashwalk tends to produce a large number of
clusters for each ground truth label, leading to very pronounced overcounting,
while DeFault and Igor generate too few clusters, leading to undercounting.
Meanwhile, \name achieves balanced groupings with low over- and
undercounting.

Looking at the results of \name for specific bug types, we see that over-
and undercounting scores are similar between bug types with the exception of
\emph{Other} bugs of the Type
\emph{Improper Restriction of Operations within the Bounds of a Memory Buffer}
(highlighted in red in \autoref{tab:eval-bug-types}),
where overcounting is significantly more prevalent than for other bug types.
The reason for this is that out of the 11 bugs of this category, 6 come from the
xmllint target where \name generates a fairly large number of clusters.
If the bugs from xmllint are excluded, this outlier effect disappears
(Over: $\mu=0$, $\sigma=0$; Under: $\mu=0.4$, $\sigma=0.89$), suggesting that
there is no clear correlation between the underlying bug type and the quality
of \name's deduplication results.
Instead, the outlier is a consequence of the specific behavior of xmllint that
we discussed in \autoref{sec:evaluation/model}.

The mean over- and undercounting scores computed for
Crashwalk vary significantly between bug types.
However, the oftentimes very large standard deviations indicate that there is
also significant variance within the bug types.
In fact, among the three bug types that show the most overcounting
(highlighted in red in \autoref{tab:eval-bug-types}), a small number of bugs
(nine out of 32) is responsible for the majority of the overcounting while the
remaining ones produce scores that are similar to those of other bug types.
This suggests that the deduplication results of Crashwalk are also not
immediately influenced by the underlying bug type.
Rather, due to the rigid nature of stack trace hashing, Crashwalk's results
exhibit a lot of overcounting for bugs that produce even marginally different
stack traces.
This will be especially prominent in cases where the same bug triggers crashes
through many different execution paths, as is the case for the aforementioned
bugs that contribute most to the overcounting.

For DeFault and Igor, over- and undercounting is similar across different bug
types, indicating that their deduplication results are not significantly
influenced by the underlying bug type either.

\subsection{Threats to Validity}

One main factor that could influence the general quality of our results is the
choice of targets in our evaluation.
Namely, all targets in our evaluation dataset are \cpp programs.
While the method of \name is applicable to data sources from programs written in
any language, we cannot ensure that the results in this case will have the same
quality as in our evaluation of \cpp targets.
Moreover, even though we used the existing Igor dataset (which is itself based
on the well-established Magma~\cite{magma} and MoonLight~\cite{moonlight}
benchmarks) and did not specifically pick the targets for our evaluation, the
quality of \name's results across a larger dataset might vary depending on the
programs and contained bugs that are under test.

Related to this is the fact that the targets in the evaluation dataset are
well-known software projects and information related to them has likely been
contained in the training sets of the \glspl{llm} that we used in our
evaluation.
This could influence the quality of \name's deduplication results for new,
unknown targets.

\section{Related Work}
\label{sec:related-work}

This section gives an overview of existing deduplication approaches and
we compare important aspects of these in \autoref{tab:comparison-matrix}.
As \name analyzes readily available crash information, such as stack traces
and \gls{asan} reports, we divide the approaches based on
whether they use similar information~\cite{crashwalk,afl-collect,honggfuzz,
    automatic_problems,tracesim,rebucket,s3m} or whether they require
more detailed information about the full program
execution~\cite{retracer,default,igor,bucketing_symbolic_analysis} or are based
on other principles~\cite{semantic_crash_bucketing,fuzzeraid}.
\begin{table}[ht]
    \centering
    \caption{Comparison of \name and other deduplication approaches.}
    \label{tab:comparison-matrix}

    \setlength{\tabcolsep}{0.1pt}
    \setlength\customtabcolsep{0.2pt}
    \rowcolors{3}{lightgray}{}
    \begin{tabular*}{\columnwidth}{l@{\extracolsep{\fill}} c@{\extracolsep{\customtabcolsep}}c@{\extracolsep{\customtabcolsep}}c@{\extracolsep{\fill}} c@{\extracolsep{\customtabcolsep}}c }
        \toprule                                                                     & \multicolumn{3}{c}{\thead{Design}}                                                                                                                  & \multicolumn{2}{c}{\thead{Implementation}}                                                         \\
        \thead{Deduplication Method}                                                 & \rotatebox{90}{\makecell[l]{Data\\Source}}  & \rotatebox{90}{\makecell[l]{Grouping\\Method}}  & \rotatebox{90}{\makecell[l]{Arbitrary\\Bug Types}}  & \rotatebox{90}{\makecell[l]{Available\\Online}}  & \rotatebox{90}{\makecell[l]{Target\\Language}}  \\
        \midrule
        \name                                                                        & Basic                                       & Cluster                                         & \checkmark                                          & \checkmark                                       & \cpp                                            \\
        DeFault~\cite{default}                                                       & ExInfo                                      & Custom                                          & \checkmark                                          & -                                                & Binary                                          \\
        FuzzerAid~\cite{fuzzeraid}                                                   & ModBin                                      & Repro                                           & \checkmark                                          & -                                                & \cpp                                            \\
        Igor~\cite{igor}                                                             & ExInfo                                      & Cluster                                         & \checkmark                                          & \checkmark                                       & Binary                                          \\
        \citeauthor{automatic_problems}~\cite{automatic_problems}                    & Basic                                       & Custom                                          & \checkmark                                          & -                                                & \cpp                                            \\
        \citeauthor{bucketing_symbolic_analysis}~\cite{bucketing_symbolic_analysis}  & ExInfo                                      & Custom                                          & \checkmark                                          & -                                                & LLVM                                            \\
        ReBucket~\cite{rebucket}                                                     & Basic                                       & Cluster                                         & \checkmark                                          & \textasteriskcentered                            & \cpp                                            \\
        RETracer~\cite{retracer}                                                     & ExInfo                                      & Custom                                          & -                                                   & -                                                & Binary                                          \\
        S3M~\cite{s3m}                                                               & Basic                                       & -                                               & \checkmark                                          & \checkmark                                       & Java                                            \\
        Trace Hashing~\cite{crashwalk,afl-collect,honggfuzz}                         & Basic                                       & Equal                                           & \checkmark                                          & \checkmark                                       & \cpp                                            \\
        TraceSim~\cite{tracesim}                                                     & Basic                                       & -                                               & \checkmark                                          & \checkmark                                       & Java                                            \\
        \citeauthor{semantic_crash_bucketing}~\cite{semantic_crash_bucketing}        & ModBin                                      & Repro                                           & -                                                   & \checkmark                                       & \cpp                                            \\
        \bottomrule
    \end{tabular*}
    \begin{minipage}{\columnwidth}
        \rowcolors{1}{}{}
        \small
        \begin{tabular*}{\columnwidth}{l@{\extracolsep{\fill}} p{0.87\columnwidth}}
            Basic                  & Ordinary crash information, \eg stack traces or \gls{asan} reports                                                              \\
            ExInfo                 & Separate, detailed information gathered during execution, \eg full execution traces or data gathered during symbolic execution  \\
            ModBin                 & Modified binary for each \gls{sci}                                                                                              \\
            Cluster                & Common clustering algorithm                                                                                                     \\
            Custom                 & Custom grouping algorithm                                                                                                       \\
            Equal                  & Check for strict equality                                                                                                       \\
            Repro                  & Use (non-)reproduction of crash with modified binaries as oracle                                                                \\
            \textasteriskcentered  & Only unofficial, incomplete implementations available
        \end{tabular*}
    \end{minipage}
\end{table}
There is also work on using \glspl{llm} to solve the related problem of
triaging and deduplicating bug reports that are written and submitted by humans
on bug tracking portals~\cite{ml_report_triage,ml_report_dedup}.
Since this problem is rather different from our fuzzing use case, we will not go
into more detail here.

\paragraph{Deduplication Using Basic Crash Information}
Stack trace-based deduplication approaches collect, for each \gls{sci}, the
stack trace at the time the program crashes and use this information to group
the \glspl{sci}.
Since analysis of these stack traces can often be done very efficiently, these
approaches tend to be quite fast.

One very simple, yet widely used way of stack trace-based deduplication is
computing hash values of stack traces and grouping together all \glspl{sci}
that produce the same hash value.
As even small changes in the stack traces lead to different hash values,
this approach often creates many groups for \glspl{sci} that have the
same underlying bug.
The results can be somewhat improved by ignoring function arguments in the
stack traces, but the tendency to produce results with low inverse purity scores
remains, as we have seen in \autoref{sec:evaluation/comparison}.
This causes additional manual work in real fuzzing workflows.
Existing tools that employ this method differ in whether they include the entire
stack trace in the computation of the hash value, like
Crashwalk~\cite{crashwalk} and afl-collect~\cite{afl-collect}, or only use
some of the topmost stack frames, such as Honggfuzz~\cite{honggfuzz}.

To improve on these hash-based approaches, other deduplication methods develop
more nuanced notions of similarity of stack traces.
\citeauthor{automatic_problems}~\cite{automatic_problems} use primitive string
metrics, such as the Levenshtein distance or the Longest Common Subsequence
algorithm, to compute similarity scores of stack traces.
An incoming crash is then assigned to the bucket that contains the trace with
the highest similarity score unless a pre-defined lower limit is not surpassed,
in which case a new bucket is created.
TraceSim~\cite{tracesim} also compares stack traces based on Levenshtein
distances but tries to improve the results by using frame weights to focus on
those frames that help the most in determining duplicate crashes.
To find good weights, they employ a suitable metric, which they optimize using
machine learning techniques.
ReBucket~\cite{rebucket} uses a simple algorithm that parses pairs of stack
traces, compares which functions occur at what positions inside them and
computes a score based on this data.
The crash bucketing is then obtained by applying a hierarchical clustering
algorithm.
In comparison to \name, these approaches extract only limited syntactic
information from the stack traces that is often prone to being disturbed.
For instance, methods that use primitive string metrics, such as the Levenshtein
distance, can yield inaccurate results if multiple distinct functions have names
that are syntactically similar, \eg \texttt{beginPass} and \texttt{beginParse}.

The stack trace analysis approach that is most similar to ours is
S3M~\cite{s3m}.
The authors propose tokenizing stack traces using a dictionary of
previously seen tokens and training a neural network that creates vector
representations of these tokenized stack traces.
Using these vectors, S3M computes for each pair of stack traces a
three-dimensional feature vector, which is then fed into another neural network
that computes a final similarity score.
In contrast to \name, S3M works with \gls{jvm} traces and does not leverage the
large amount of information that state-of-the-art \glspl{llm} learn during their
training.
Moreover, their use of a second neural network for computing similarity scores
from feature vectors makes S3M less transparent and they do not propose a
method that groups the \glspl{sci} based on those similarity scores.

\paragraph{Other Deduplication Approaches}
There are also deduplication methods that do not (solely) analyze stack traces
or other basic crash data.
Often, these approaches monitor or manipulate the execution of each \gls{sci}
to gather additional information that is not captured by ordinary stack traces
and use it to produce deduplication results.

RETracer~\cite{retracer} performs binary analysis based on memory dumps to find
the root cause of crashes.
However, it is limited in the types of crashes it can analyze and its target use
case is crash triage based on limited information that is collected on end user
machines rather than active software testing through fuzzing.
\citeauthor{bucketing_symbolic_analysis}~\cite{bucketing_symbolic_analysis} use
symbolic execution to capture semantic information of different execution paths,
which is then used to group \glspl{sci}.
While \name focuses on deduplicating crashes returned by a fuzzing campaign,
\citeauthor{bucketing_symbolic_analysis} aim to assist developers in debugging
software by collecting information about the semantics of the program.
Furthermore, being based on symbolic execution, their method is prone to
long run times, especially for large real-world programs.
The method developed by
\citeauthor{semantic_crash_bucketing}~\cite{semantic_crash_bucketing}
creates approximate bug fixes and groups \glspl{sci} together if they can be
caught by the same approximate bug fix.
This approach is very limited in the types of bugs it can handle since the
required approximate bug fixes strongly depend on the bug type and need to be
written manually.
This is incompatible with the real-world deduplication use case that \name
targets.
FuzzerAid~\cite{fuzzeraid} extracts for each \gls{sci} a fault signature, \ie a
small selection of statements from the original program that is necessary to
reproduce the crash.
If one input leads to a crash of another input's fault signature, then these two
inputs are grouped together.
Unfortunately, we cannot compare \name's deduplication capabilities to
FuzzerAid's, as there is no source code of FuzzerAid publicly available and the
evaluation in the paper is based on a different dataset than ours.

DeFault~\cite{default} quantifies the relevance of basic blocks of execution
traces to crashing behavior by computing mutual information scores.
These are a measure of how likely an execution trace is to correspond to a crash
given that it contains a specific basic block.
To do this, they also include non-crashing traces in their analysis.
The authors then devise an algorithm that groups \glspl{sci} based on the
mutual information scores of their associated stack traces.
Since there is no source code available for DeFault, we implemented a
prototype implementation based on the paper's description and the evaluation
showed that it cannot produce results that are as accurate as those of \name.
Igor~\cite{igor} clusters \glspl{sci} based on the similarity of control flow
graphs.
To improve the deduplication quality, Igor does not use the
control flow graphs that are directly associated to the \glspl{sci} but
instead introduces a minimum-coverage fuzzing phase that finds another \gls{sci}
that still triggers the same bug but does so via an execution path that is
as short as possible.
As we have seen in \autoref{sec:evaluation/comparison}, Igor produces
deduplication results that are not as accurate as those of \name and introduces
complexity as well as long run times when integrated into real fuzzing workflows.

\section{Conclusion and Future Work}
\label{sec:conclusion}

The large amounts of data collected during fuzzing campaigns
can quickly become overwhelming and waste valuable human resources.
To remedy this, we presented and implemented \name, an approach that uses
\gls{llm} embeddings of crashing data to deduplicate \glspl{sci}.
We evaluated \name on a large dataset and have seen that it produces more
accurate deduplication results than stack trace hashing approaches and even more
complex state-of-the-art deduplication methods that require longer run times.
To enable reproducibility of our results and facilitate future
research, we make our \name implementation and artifacts publicly
available.

There are two directions of future work that might enable improvements of \name.
First, it would be interesting to investigate what further data sources beyond
stack traces and \gls{asan} reports carry meaningful information about crashing
behavior.
Thanks to the flexible design of \name, these could be integrated into \name and
enable better deduplication for targets that prove challenging when relying
solely on stack traces and \gls{asan} reports.
Secondly, while the general \glspl{llm} that we evaluated already produce
embeddings that yield good deduplication results, it would be interesting to
investigate whether specifically fine-tuning \glspl{llm} on the task of
deduplication leads to even more accurate bucketings.

\begin{acks}
  We would like to thank the anonymous reviewers for their valuable suggestions
  for improving this paper.
  This research was supported by the
  \grantsponsor{StMWi}{Bavarian Ministry of Economic Affairs, Regional Development and Energy}{}.
\end{acks}

\setcitestyle{numbers, sort&compress}
\bibliographystyle{ACM-Reference-Format}

\bibliography{references}

\end{document}